\documentstyle[seceq,mbf,preprint]{ptptex}
 \notypesetlogo
  \def\bea*{\begin{eqnarray*}}
 \def\eea*{\end{eqnarray*}}
\def\ind{\indent}
 \def\nn{\nonumber}
 \def\be{\begin{equation}}
 \def\ee{\end{equation}}
 \def\hx{{\hat x}}
 \def\hS{{\hat S}}
 \def\btheta{{\bar \theta}}
 \def\htheta{{\hat \theta}}
 \def\hvarphi{{\hat \varphi}}
 \def\beq{\begin{eqnarray}}
 \def\eeq{\end{eqnarray}}
 \def\ba{\begin{array}}
 \def\ea{\end{array}}
 \def\cL{{\mbox {${\cal L}$}}}
 \def\cM{{\mbox {${\cal M}$}}}

 \def\bl{{\mbox {{\mbf l}}}}
 \def\bk{{\mbox {{\mbf k}}}}
  \def\bq{{\mbox {{\mbf q}}}}

 \def\bp{{\mbox {{\mbf p}}}}
 \def\cL{{\mbox {${\cal L}$}}}

  \def\lslash{l{\raise 0.8pt\hbox{$\!\!\!/$}}}
  \def\qslash{q{\raise 0.8pt\hbox{$\!\!\!/$}}}
  \def\pslash{p{\raise 0.8pt\hbox{$\!\!\!/$}}}
\title{%
A New Gauge-Invariant Regularization Scheme Based on\\
Lorentz-Invariant Noncommutative Quantum Field Theory
}
\vspace{5mm}
\author{%
  Katsusada {\sc Morita}
}
\inst{%
{\it  Department of Physics, 
  Nagoya University, Nagoya 464-8602, Japan}
}
\vspace{5mm}
\abst{%
The IR/UV mixing
in the non-commutative (NC) field theory
is investigated in Carlson-Carone-Zobin (CCZ) formalism of
Lorentz-invariant NC field theory
provided that
the fields 
are `independent' of the `internal' coordinates $\theta^{\mu\nu}$.
A new regularization scheme called
NC regularization is 
then proposed, which removes 
the Lorentz-invariant 
IR singularity by subtraction
from the theory.
It
requires 
the usual UV limit $\Lambda\to \infty$
to be accompanied with the commutative
limit $a\to 0$ with $\Lambda^2a^2$ fixed, 
where $a$ is the length
parameter in the theory. 
The new UV limit 
gives
the usual renormalized amplitude of the 
one-loop self-energy diagram of $\phi^3$ model.
It is shown that
the new regularization is gauge-invariant, that is,
the non-transverse part of the vacuum polarization
in QED
is automatically transverse in Lorentz-invariant NCQED 
but the two transverse pieces,
one of which is already transverse in QED,
possesses Lorentz-invariant IR singularity
which should be `subtracted off' at
zero external momentum squared.
The subtraction leads to the same result as
the renormalized one by Pauli-Villars or dimensional regularizations.
Other diagrams with three-point vertices
which contribute to the photon self-energy in
Lorentz-non-invariant NCQED all vanish due to
Lorentz invariance under the assumption adopted, 
while the tadpole diagram
gives a finite contribution
to the charge renormalization,
which vanishes if $\Lambda^2a^2\to 0$.
Lorentz-invariant
NC $\phi^4$ and scalar Yukawa models
are also discussed in the one-loop approximation.
A comment is made that
Lorentz invariance  
might lead to a decoupling of $U(1)$
from $SU(N)$ in NC $U(N)$ gauge theory.
}
\begin{document}
\maketitle
\baselineskip=17.8pt
%
\section{Introduction}
It is well-known that
noncommutative (NC) quantum field theory (NCQFT) 
is annoyed with
IR singularity.
The IR/UV mixing\cite{1)}
means that
the nonplanar diagram is made UV finite
due to Moyal phase depending
on the loop momenta,\cite{2)}
while it
is UV divergent in the commutative limit.
It is
deeply rooted in the noncommutativity assumption.
It should be recalled, however, that
the IR singularity is not Lorentz-invariant 
because it depends on the constant
noncommutativity parameter explicitly.
It should also be mentioned that
NC gauge theory (NCGT) introduces many new vertices due to 
the noncommutativity so that
the elimination of the IR singularity, 
if possible, becomes very cumbersome.
\\
\ind
In the literature there
have been proposed\cite{1),3)} several interpretations
of the IR/UV mixing in NCQFT with Lorentz violation.
In this paper we propose quite a new
interpretation.
It is only a Lorentz-invariant
formulation of the IR/UV mixing
that one could understand a physical origin
of it, thereby evading the astonishing phenomenon 
consistently.
This viewpoint comes from our belief
that the Lorentz violation
in NCQFT invalidates the
correspondence principle 
that the commutative limit of
NCQFT must be reduced
to quantum field theory (QFT) in a {\it smooth} way.
This single requirement
excludes as a consistent NCQFT
all Lorentz-non-invariant
NCQFTs which fail to control 
a singular behavior in the commutative limit.
\\
\ind
From the correspondence principle
it seems to be essential to reformulate
NCQFT in a Lorentz-invariant way
so that the IR singularity
occurring in any Feynman amplitude
is identified unambiguously so that
it is `subtracted off'
uniquely up to arbitrary subtraction point.
In fact, the new gauge-invariant
regularization scheme we propose in this paper is
based on Carlson-Carone-Zobin (CCZ)\cite{4)} formalism of
Lorentz-invariant NCQFT.
We call the new regularization NC regularization. 
The NC regularization is
related to the elimination of the 
IR singularity in CCZ formalism
which predicts Lorentz-invariant,
finite amplitude off the IR region
(a point in Euclidean metric).
\\
\ind
CCZ formalism\cite{4)}
makes use of the 
Doplicher, Fredenhagen and Roberts (DFR)
algebra.\cite{5)}
In Ref.~6) the persistence of the IR singularity
in CCZ formalism of NC $\phi^4$ model
was
proved in Euclidean metric
using Gaussian weight function
and it was
pointed out that
the IR
singularity in NC $\phi^4$ model
becomes Lorentz-invariant \footnote{This
means that the IR singularity no longer depends on
the noncommutativity parameter
explicitly but depends on it only through
Lorentz-invariant moments in the $\theta$-space.}
due to
the presence of invariant
damping factor
instead of the oscillating Moyal
phase. The invariant damping factor
was obtained through $\theta$-integration
in CCZ formalism.
To avoid the Lorentz-invariant IR singularity
a new UV limit was then proposed such that
\begin{eqnarray}
\Lambda^2&\to&\infty,\qquad
a^2\to 0,\qquad
\Lambda^2 a^2:{\rm fixed},
\label{eqn:1-1}
\end{eqnarray}
where $\Lambda$ is UV cutoff
and $a$ is a length parameter
in the theory:
\begin{eqnarray}
\theta^{\mu\nu}=a^2{\bar\theta}^{\mu\nu}
\label{eqn:1-2}
\end{eqnarray}
with ${\bar\theta}^{\mu\nu}$
dimensionless. Here,
the noncommutativity parameter
$\theta^{\mu\nu}$ is an antisymmetric 
$c$-number tensor characterizing an irreducible
representation of the DFR algebra
which is discussed in the next section
in more details. The usual
UV limit $\Lambda\to\infty$
should not be naively applicable in Lorentz-invariant
NCQFT
which involves a length parameter whose inverse
should supply a UV cutoff. Then the question
of order between the UV limit and
the commutative limit $a\to 0$,
which leads to the IR/UV mixing, would become meaningless.
The two limits should be taken simultaneously.
The real significance of the new UV limit was not
understood, however, in Ref.~6) because 
the analytic continuation back to
Minkowski momenta was not attempted with
correct subtraction.
We would like to reveal
it in the present paper.
The conclusion is that the Lorentz-invariant IR singularity
in CCZ formalism with
the additional assumption that
the fields (even subject to
$\ast$-gauge transformation)
are `independent' of the `internal' coordinates $\theta^{\mu\nu}$,
can be
eliminated by introducing
a new
regularization with arbitrary subtraction point.
The new regularization turns out to be gauge-invariant
because the non-transverse part of the vacuum polarization
tensor
in QED becomes automatically transverse in
the Lorentz-invariant NCQED.
The subtraction
of the IR singularity then
reproduces the well-known
renormalized amplitude
obtained via Pauli-Villars or
dimensional regularizations.
It is interesting to
realize that all couplings
with three point-vertices
in Lorentz-non-invariant NCQED\cite{7)}
vanish due to Lorentz invariance
under the additional assumption alluded to above,
whereas the remaining tadpole
diagram is shown to give rise to
a finite charge renormalization
which vanishes if $\Lambda^2a^2\to 0$.
In this connection tadpole diagram
in Lorentz-invariant NC $\phi^4$ model
and the fermion loop with Lorentz-invariant NC scalar 
Yukawa model
are also discussed.
\\
\ind
This paper is organized as follows.
In the next section we succinctly review CCZ formalism.
It is pointed out in \S3 that
the unitarity problem\cite{8)} in NC $\phi^3$
model is avoided in the new UV limit.
Using the Gaussian weight function
we show 
in \S4 that the IR singularity
in Lorentz-invariant NC $\phi^3$
model
extends over the external
non-spacelike momenta for the one-loop
self-energy
diagram. This demands
an introduction of the UV cutoff
in the Schwinger integration.
The precise form of the IR singularity
depends of course on the form of
the weight
function, but 
the presence of the IR singularity in CCZ formalism
is a general phenomenon.
The new UV limit (\ref{eqn:1-1})
to remove the IR divergence
renormalizes the one-loop self-energy
diagram by subtraction.
The subtraction point can be chosen arbitrary.
Vacuum polarization in the Lorentz-invariant NCQED
is shown in \S5 to be automatically transverse
and the new UV limit (\ref{eqn:1-1})
with the correct subtraction to
avoid IR singularity
reproduces the well-known renormalized amplitude
(in the one-loop approximation).
It is shown in \S6 that
the diagrams with three-point vertices
which contribute to the photon self-energy in
Lorentz-non-invariant QED\cite{7)} all vanish due to
Lorentz invariance
(using the action (\ref{eqn:2-11}) in the next section), 
while the tadpole diagram
leads to a finite charge
renormalization
which vanishes if $\Lambda^2a^2\to 0$.
The tadpole diagram in Lorentz-invariant
NC $\phi^4$ model and the
fermion loop with Lorentz-invariant
NC scalar Yukawa model are
also considered in \S6.
The last section contains a short comment on
Lorentz-invariant NC $U(N)$ gauge theory 
and discusses further problems.
\section{CCZ formalism of Lorentz-invariant NCQFT} 
NC field theory is formulated
based on the $\theta$-algebra
\be
[\hx^\mu,\hx^\nu]=i\theta^{\mu\nu},\quad
\quad\mu,\nu=0,1,2,3,
\label{eqn:2-1}
\ee
where the space-time coordinates are
represented by hermitian operators $\hx^\mu$
with $(\theta^{\mu\nu})$ being a real antisymmetric 
constant matrix. 
Any field in NC field theory is an operator-valued function, $\hvarphi(\hx)$.
In terms of 
the Weyl symbol $\varphi(x)$ defined through
\beq
\hvarphi(\hx)&=&\displaystyle{1\over (2\pi)^4}\int\!d^4kd^4x\varphi(x)
e^{-ikx}e^{ik{\hat x}},
\label{eqn:2-2}
\eeq
with $k\hx\equiv k_\mu\hx^\mu$,
NC field theory becomes
a nonlocal
field theory on the ordinary space-time with
the point-wise multiplication of the field variables
being replaced by the Moyal $\ast$-product
corresponding to the product of the operators,
\begin{eqnarray}
\hvarphi_1(\hx)\hvarphi_2(\hx)\longleftrightarrow
\varphi_1(x)*\varphi_2(x)\equiv
\varphi_1(x)e^{\frac i2
\theta^{\mu\nu}{\mbox{\scriptsize$
{\overleftarrow{{\partial_\mu}}}$}}
{\mbox{\scriptsize$\overrightarrow{\partial_\nu}$}}}
\varphi_2(x).
\label{eqn:2-3}
\end{eqnarray}
The action defining NC field theory is then given by
\be
S={\rm tr}[{\hat{\cL}}(\hvarphi(\hx), \partial_\mu\hvarphi(\hx))]
=\int\!d^{\,4}x\cL(\varphi(x), \partial_\mu\varphi(x))_*,
\label{eqn:2-4}
\ee
where we have normalized tr$e^{ik\hx}=(2\pi)^4\delta^4(k)$
and the subscript $\ast$ of the Lagrangian indicates that
the $\ast$-product should be taken for all products
of the field variables. 
\\
\ind
NCQFT is defined as a QFT based on the
classical action (\ref{eqn:2-4}).
Feynman rules are derived
using the path integral.
In addition to the fact that
the nonlocality of the interactions
leads to the IR/UV mixing,
we face to the problem of Lorentz violation
in NCQFT
defined on the $\theta$-algebra.
However, Lorentz invariance is one of the most fundamental
symmetries in QFT.
Even if the space-time
were not a continuum but were
instead described by a noncommutative geometry, say, at the
Planck scale,
Lorentz invariance should be maintained
because more but not less symmetries are expected
to be effective at shorter distances.
To retrospect
a Lorentz-invariant NC space-time
was first considered by Snyder\cite{9)}
(see also Yang\cite{10)})
but it is absolutely unrelated with
NCQFT based on the
$\theta$-algebra (\ref{eqn:2-1}).
On the contrary,
DFR\cite{5)}
defined quantum space based on the Lorentz-covariant
algebra
\beq
[{\hat x}^\mu,{\hat x}^\nu]&=&i{\hat\theta}^{\mu\nu},\;\;\;
[{\hat\theta}^{\mu\nu},{\hat x}^\nu]=0=
[{\hat\theta}^{\mu\nu},{\hat\theta}^{\rho\sigma}],
\quad\mu,\nu,\rho,\sigma=0,1,2,3,
\label{eqn:2-5}
\eeq
where $\htheta^{\mu\nu}$ is an antisymmetric second-rank
tensor operator, and set up a Lorentz-invariant
NCQFT on it with some constraint
on the non-commutativity
parameter $\theta^{\mu\nu}$, the eigenvalue of the operator
$\htheta^{\mu\nu}$.
Feynman rules of the theory
was formulated by Filk\cite{2)}
who
considered a single irreducible representation
of the DFR algebra (\ref{eqn:2-5}),
which essentially is tantamount
to restricting to the $\theta$-algebra (\ref{eqn:2-1}),
and found that, although planar diagrams
are still divergent as in QFT,
the noncommutativity renders nonplanar diagrams 
convergent.
Since the revival of NCQFT by the
refinement of Seiberg and Witten\cite{11)}
connected to the string theory,
many researchers studied various aspects of
NCQFT based on the $\theta$-algebra.
Main results obtained in the first stage
of the extensive study
are the IR/UV mixing\cite{1)} and
the charge quantization.\cite{7)}
\\
\ind
In the last year CCZ\cite{4)}
successfully constructed NCGT without Lorentz violation
by `contracting' Snyder's algebra
to obtain the DFR algebra.
They also asserted that all irreducible representations
of the DFR algebra
should be taken into account
because $\theta^{\mu\nu}$ plays a role of `internal'
coordinate in their formulation.
Let us now summarize CCZ formalism.
\\
\ind
Any field defined on
the DFR algebra
is the operator, $\varphi(\hx,\htheta)$.
The associated Weyl symbol
now depends on the eigenvalue $\theta^{\mu\nu}$ of 
the operator $\htheta^{\mu\nu}$, written as $\varphi(x,\theta)$,
and the new correspondence is given by
\begin{eqnarray}
&&\hvarphi(\hx,\htheta)=
\int\!d^{\,4}kd^{\,6}\sigma
{\tilde\varphi}(k,\sigma)e^{ik\hx+i\sigma\htheta}\longleftrightarrow
\varphi(x,\theta)=
\int\!d^{\,4}kd^{\,6}\sigma{\tilde\varphi}(k,\sigma)e^{ikx+i\sigma\theta}\nn\\[2mm]
&&\hvarphi_1(\hx,\htheta)\hvarphi_2(\hx,\htheta)\longleftrightarrow
\varphi_1(x,\theta)*\varphi_2(x,\theta)
\equiv
\varphi_1(x,\theta)e^{\frac i2
\theta^{\mu\nu}{\mbox{\scriptsize$
{\overleftarrow{{\partial_\mu}}}$}}
{\mbox{\scriptsize$\overrightarrow{\partial_\nu}$}}}
\varphi_2(x,\theta),
\label{eqn:2-6}
\end{eqnarray}
where $\sigma\htheta\equiv\frac 12\sigma_{\mu\nu}\htheta^{\mu\nu}$.
To be more precise
the Moyal $*$-product corresponding to the operator
product $\varphi_1(\hx,\htheta)\varphi_2(\hx,\htheta)$
is given by
\begin{eqnarray}
&&\displaystyle{1\over (2\pi)^{10}}
\int\!d^4kd^6\sigma e^{ikx+i\sigma\theta}
{\rm tr}[\varphi_1(\hx,\htheta)
\varphi_2(\hx,\htheta)
e^{-ik{\hat x}-i\sigma\htheta}]\nn\\[2mm]
&&\quad\quad=
W(\theta)\varphi_1(x,\theta)*\varphi_2(x,\theta)
=W(\theta)e^{\frac i2
\partial_1\wedge \partial_2}\varphi_1(x_1,\theta)
\varphi_2(x_2,\theta){\big|}_{x_1=x_2=x},
\label{eqn:2-7}
\eeq
where $\sigma\theta\equiv \frac 12\sigma_{\mu\nu}\theta^{\mu\nu}$,
$\partial_1\wedge \partial_2=\theta^{\mu\nu}
\partial_{1,\mu}\partial_{2,\nu}$
and we have normalized tr$e^{i\sigma\htheta}=
{\tilde W}(\sigma)$, which is the
Fourier component of $W(\theta)$,
\begin{eqnarray}
W(\theta)=\frac 1{(2\pi)^6}
\int\!d^6\sigma {\tilde W}(\sigma)e^{-i\sigma\theta},\;\;\;
{\tilde W}(0)=1.
\label{eqn:2-8}
\eeq
Because of the extra `internal' variable,
one needs an integration over the extra 6-dimensional
variable $\theta^{\mu\nu}$.
Integrating (\ref{eqn:2-7}) over $x$ and
$\theta$ yields
the formula
\begin{eqnarray}
{\rm tr}[\varphi_1(\hx,\htheta)
\varphi_2(\hx,\htheta)
]
=\int\!d^4xd^6\theta
W(\theta)\varphi_1(x,\theta)*\varphi_2(x,\theta).
\label{eqn:2-9}
\eeq
The action (\ref{eqn:2-4})
is thus replaced with
\be
S_{\rm CCZ}={\rm tr}[{\hat{\cL}}(\hvarphi(\hx,\htheta), 
\partial_\mu\hvarphi(\hx,\htheta))]
=\int\!d^{\,4}xd^{\,6}\theta\,
W(\theta)\cL(\varphi(x,\theta), \partial_\mu\varphi(x,\theta))_*.
\label{eqn:2-10}
\ee
This form of the Lorentz-invariant NCQFT action with 
Lorentz-invariant, normalized weight function $W(\theta)$
was first obtained by CCZ.\cite{4)}
In this paper we assume that fields even subject to
$\ast$-gauge transformation are all `independent' of the
`internal' coordinate, $\theta^{\mu\nu}$,
so that we keep away from
the quantization problem of the field
$\varphi(x,\theta)$ on NC space-time.
That is, in what follows,
we assume the Lorentz-invariant action
\be
\hS={\rm tr}[{\hat{\cL}}(\hvarphi(\hx), 
\partial_\mu\hvarphi(\hx))]
=\int\!d^{\,4}xd^{\,6}\theta\,
W(\theta)\cL(\varphi(x), \partial_\mu\varphi(x))_*.
\label{eqn:2-11}
\ee
Consequently,
the only difference from the action
(\ref{eqn:2-4}) lies in the $\theta$-integration.
\footnote{CCZ\cite{4)} argued that
for a theory without gauge invariance,
one may simply choose $\varphi(x,\theta)=\varphi(x)$,
while such a choice is no longer possible for a gauge-invariant theory.
On the other hand, even propagators
cannot be simply obtained from (\ref{eqn:2-10})
if we stick to the `explicit' $\theta$-dependence
of fields for gauge theory without recourse to
the $\theta$-expansion. To dispense with such
a formidable quantization problem we employ
the action (\ref{eqn:2-11}) for both non-gauge and gauge theories
in this paper.
Nonetheless, the field $\varphi(x)$ 
subject to $\ast$-gauge transformation
is still defined
on NC space-time and
can be
$\theta$-expanded as in Ref.~4) using 
Seiberg-Witten map\cite{11)} as advocated by Jurc{$\breve{\rm o}$}
et al.\cite{12)}
This is reflected by the fact that
the action (\ref{eqn:2-11}) is simply
obtained by integrating the action (\ref{eqn:2-4}) over $\theta$,
whereas the $\theta$-expansion in Ref. 4)
is no different from the approach\cite{12)}
based on the action (\ref{eqn:2-4}) except for the
$\theta$-integration. This
diminishes a role of $\theta^{\mu\nu}$
as an `internal' coordinate.}
This difference leads, however, to a nontrivial modification
in the vertex structure. For instance,
the three point vertex proportional to
sin factor as occurs in Lorentz-non-invariant
QED all vanish because the weight function
is odd by Lorentz invariance.
Although this conclusion, which is applicable also to 
Lorentz-invariant NC non-Abelian gauge theory,
does not depend on
a particular form of the weight function,
we have to assume a concrete form
of the weight function
to identify the IR singularity in a definite way.
In this paper we assume a Gaussian weight function
as in Ref.~6).
\section{IR singularity and unitarity
problem in NC $\phi^3$ model} 
Gomis and Mehen\cite{8)}
pointed out a breakdown of the
unitarity relation
in space-time noncommutative QFT,
while they claimed that the unitarity
relation in space-space noncommutative QFT
holds true in a form involving
the noncommutativity parameter explicitly.
The present section is devoted to
a discussion on the
unitarity problem in relation to the
IR/UV mixing \footnote{The 
interplay between the 
IR/UV mixing
and the unitarity problem was also discussed by
Chu et al.\cite{13)}
in a different context. }
and Lorentz invariance.
The conclusion will underlie the
philosophy of the present paper.
\\
\ind
To this purpose let us consider NC $\phi^3$ model,
\beq
S&=&\int\!d^4x[\frac 12\partial_\mu\phi(x)*
\partial^\mu\phi(x)-\frac 12m^2\phi(x)*\phi(x)
-\frac {\lambda}{3!}\phi(x)*
\phi(x)*\phi(x)],
\label{eqn:3-1}
\eeq
where $\phi(x)$ is a scalar field, $m$ is 
the mass parameter and $\lambda$ is the
coupling constant.
The one-loop amplitude for the self-energy diagram \footnote{In
this paper the subscript 2 of the amplitude
indicating
the second-order approximation
is neglected for simplicity.}
is given by
\begin{eqnarray}
iM
&=&\frac{\lambda^2}4\int\!\frac{d^4l}{(2\pi)^4}
\frac{1+\cos{(p\wedge l)}}{((p-l)^2-m^2+i\epsilon)(l^2-m^2+i\epsilon)},
\label{eqn:3-2}
\end{eqnarray}
where $p$ is the external momentum, 
and $p\wedge l=p_\mu\theta^{\mu\nu}l_\nu$.
The real part of the Moyal phase
$e^{\frac i2p\wedge l}$ is associated with each vertex,
hence the product
$\cos{(\frac 12p\wedge l)}\cos{(\frac 12p\wedge l)}=
\frac{1+\cos{(p\wedge l)}}2$ determines the extra NC factor.
The part without $\cos{(p\wedge l)}$
is called the planar diagram,
while
the part with $\cos{(p\wedge l)}$
the nonplanar diagram.\cite{1),3)}
Considering them together and
using Feynman parameter and Schwinger representation
we write (\ref{eqn:3-2})
as a convergent integral
\beq
M&=&\frac{\lambda^2}{64\pi^2}
\int_0^1\!dx\int_0^\infty\!dss^{-1}
\big(e^{-s(\Delta(p^2,m^2)-i\epsilon)}+
e^{-s(\Delta(p^2,m^2)-i\epsilon)-\frac {p\circ p}{4s}}\big)
e^{-\frac 1{s\Lambda^2}},
\label{eqn:3-3}
\eeq
with $\Delta(p^2,m^2)=-p^2x(1-x)+m^2$,
where
we define
\beq
p\circ p&\equiv&p^\mu\theta^2_{\mu\nu}p^\nu
=(p^0\theta_{0i})^2-(p^i\theta_{i0})^2
-2p^0\theta_{0i}\theta_{ij}p^j
+(p_i\theta^{ij})^2,
\label{eqn:3-4}
\eeq
and the last factor $e^{-\frac 1{s\Lambda^2}}$
in (\ref{eqn:3-3})
is for the regularization.\footnote{The 
convergence of the integral (\ref{eqn:3-3})
limits $p^2<4m^2$ and $p\circ p/4+\frac 1{\Lambda^2}>0$.}
It is important to remember that
$M$ depends on $p\circ p$ as well as $p^2,
M=M(p^2, p\circ p,\Lambda^2)$.
The Lorentz violation in the conventional
NC field theory is manifested in the fact that
neither $p\circ p$ nor the amplitude $M$ are Lorentz-invariant.
\\
\ind
It is straightforward to obtain from (\ref{eqn:3-3})
\beq
M&=&\frac{\lambda^2}{32\pi^2}\int_0^1\!dx
[K_0\big(2\sqrt{\Delta(p^2,m^2)/\Lambda^2}\big)
+K_0\big(2\sqrt{(p\circ p/4+1/\Lambda^2)\Delta(p^2,m^2)}\big)],
\label{eqn:3-5}
\eeq
where $\Delta(p^2,m^2)$ is assumed to be positive
and $K_0$ is the modified Bessel function
of the second kind.
If we first take the UV limit $\Lambda^2\to\infty$
for the nonplanar diagram
represented by the
second term in (\ref{eqn:3-5}),
it
is finite for $p\circ p>0$
but shows a singular behavior in the
IR (commutative) limit $p\circ p\to 0$. 
The singular behavior is called the IR singularity.
If, on the other hand,
we first take the commutative limit $p\circ p\to 0$
and then the UV limit,
the nonplanar diagram is log divergent as for the planar one.
Consequently, the UV limit and the commutative
limit are not commutative
for the nonplanar diagram. This is called
the IR/UV mixing which occurs only for the nonplanar
diagrams.
It is easy to show that, by first taking the UV limit
with fixed $p\circ p>0$,
(\ref{eqn:3-5}) behaves like 
\beq
M&=&-\frac{\lambda^2}{32\pi^2}\int_0^1\!dx
[\big(\gamma+\ln{(\sqrt{\Delta(p^2,m^2)/\Lambda^2})}\big)\nn\\[2mm]
&&+I_0((p\circ p)\Delta(p^2,m^2))
(\gamma +\ln{\big(\frac {\sqrt{(p\circ p)\Delta(p^2,m^2)}}2\big)})],
\label{eqn:3-6}
\eeq
where $I_0$ is
the modified Bessel function
of the first kind and $\gamma$ is Euler's constant.
The first term exhibits the log divergence
of the planar diagram, 
which is to be subtracted off in any way,
while
the second one is finite as far as $p\circ p$
is positive but becomes singular
as $p\circ p\to 0$.
Due to the presence of the IR singularity
which is a branch point in this model,
it develops an imaginary part
when $p\circ p$ becomes negative
which can occur for space-like $p$ if $\theta^{0i}\ne 0$.
This is the reason that the authors in Ref. 8)
claimed the unitarity violation in space-time
noncommutative QFT.\footnote{Gomis and Mehen
confined only to the nonplanar diagram
and did not introduce the UV cutoff parameter.
Their answer is given by the second term
of (\ref{eqn:3-5}) without $1/\Lambda^2$.}
On the contrary to their assertion,\cite{8)}
there may not be unitarity violation
if we take the view that
{\it the IR singularity should be `subtracted off'
on physical grounds}
just as the UV divergence is subtracted off
to get finite result in QFT and
the `subtraction' would eliminate
the IR singularity, namely, the branch point at
$p\circ p=0$.
In other words,
there might be a regularization method
to remove the IR singularity from the theory.
Such a regularization, if exists,
should be associated with
a regularization of the UV divergence because
the IR singularity puts in an appearance
by the very existence of the UV divergence
in the commutative amplitude.
In this sense
the unitarity problem
is intimately related to the IR/UV mixing.
Note, however, that the limit $p\circ p\to 0$
has no invariant meaning.
Hence, elimination
of the singularity at $p\circ p=0$
is insufficient to reguralize the UV divergence
in an invariant way.
This is clear from the fact that
the IR/UV mixing occurs only for the nonplanar
diagrams.
\\
\ind
Before defining such a regularization method
which requires a Lorentz-invariant formulation of the IR/UV mixing,
we would like to point out that
the unitarity relation
obtained by Gomis and Mehen\cite{8)}
for time-like momentum contains an
inconsistency.
Using the relation
Im$2K_0(e^{-i\pi/2}z) 
=\pi J_0(z)$ for real positive
$z$, where $J_0$ is the Bessel function,
(\ref{eqn:3-5}) gives the imaginary part for $p^2>4m^2$
\beq
{\rm Im}\,M&=&\frac{\lambda^2}{64\pi}\int_{\frac{1-\rho}2}
^{\frac{1+\rho}2}\!dx
[J_0\big(\sqrt{(-\Delta(p^2))/\Lambda^2}\big)
+J_0\big(\sqrt{(p\circ p+1/\Lambda^2)(-\Delta(p^2))}\big)],
\label{eqn:3-7}
\eeq
where $\rho=\sqrt{1-\frac{4m^2}{p^2}}$.
In the commutative theory
it is well-known that the imaginary part is regularization-independent.
If the same is true also in NCQFT,
Gomis-Mehen's result\cite{8)}
is recovered
by simply taking the limit $\Lambda^2\to \infty$
in (\ref{eqn:3-7})
\beq
{\rm Im}\,M&=&\frac{\lambda^2}{64\pi}\rho
+\frac{\lambda^2}{64\pi}\int_{\frac{1-\rho}2}
^{\frac{1+\rho}2}\!dxJ_0\big(\sqrt{p\circ p(-\Delta(p^2))}\big)\nn\\[2mm]
&=&\frac{\lambda^2}{64\pi}\rho
+\frac{\lambda^2}{32\pi}
\frac{\sin{(\rho\sqrt{p^2p\circ p}/2)}}{\sqrt{p^2p\circ p}},
\label{eqn:3-8}
\eeq
where we have assumed $p\circ p>0$. 
This result
can also be obtained directly from (\ref{eqn:3-6}).
\\
\ind
On the other hand,
the unitarity sum 
derived from the cutting rule
in the same approximation
is evaluated\cite{8)} to be
for $p^2>4m^2$
\beq
\sum|M|^2&=&
\frac {\lambda^2}{2(2\pi)^2}
\int\!\frac{d^3{\mbf k}}{2\omega_{\mbf k}}
\int\!\frac{d^3{\mbf q}}{2\omega_{\mbf q}}
\delta^4(p-q-k)\frac {1+\cos{(p\wedge k)}}2\nn\\[2mm]
&=&
\frac{\lambda^2}{4\cdot 32\pi^2}\rho
\int\!d\Omega[1+\cos{(p\wedge k)}]
=\frac{\lambda^2}{32\pi}\rho
+\frac{\lambda^2}{16\pi}
\frac{\sin{(\rho\sqrt{p^2p\circ p}/2)}}{\sqrt{p^2p\circ p}}.
\label{eqn:3-9}
\eeq
This seems to verify the unitarity relation 2Im$M=\sum|M|^2$
for time-like momentum in NCQFT.
\\
\ind
It should be remarked, however,
that, in obtaining this result,
the integral
\beq
I&=&\int\!d\Omega\cos{(p\wedge k)}
\label{eqn:3-10}
\eeq 
has to be calculated.
An almost trivial method,
which might be employed by many authors,
is presented for completeness.
For time-like $p$ we go over to
the rest frame, $p^0\ne 0, \bp=0$:
\beq
p\wedge k&=&p_0\theta^{0i}k_i
={\tilde {\mbf p}}\cdot\bk=|{\tilde {\mbf p}}||\bk|
\cos{\theta},
\label{eqn:3-11}
\eeq
where $\theta$ is the angle between $\bk$ and
\beq
{\tilde {\mbf p}}=
({\tilde p}_1=\theta^{01}p_0,
{\tilde p}_2=\theta^{02}p_0,
{\tilde p}_3=\theta^{03}p_0).
\label{eqn:3-12}
\eeq
It follows from
(\ref{eqn:3-4}) that
\beq
p\circ p&=&|{\tilde{\mbf p}}|^2
\label{eqn:3-13}
\eeq
in the rest frame. Consequently, we have
\beq
p\wedge k&=&\sqrt{p\circ p}|\bk_{\rm cm}|\cos{\theta},
\quad |\bk_{\rm cm}|=\sqrt{\frac{p^2}4-m^2}.
\label{eqn:3-14}
\eeq
Thus we arrive at the result
\beq
I&=&2\pi\int_0^\pi\!d\theta\sin{\theta}\cos{(\sqrt{p\circ p}|\bk_{\rm cm}|\cos{\theta})}
=4\pi
\frac{\sin{(\sqrt{p\circ p}|\bk_{\rm cm}|)}}{\sqrt{p\circ p}|\bk_{\rm cm}|}.
\label{eqn:3-15}
\eeq
This gives (\ref{eqn:3-9}). 
There seems nothing wrong in this proof of
the unitarity relation.
However,
we have only checked the unitarity relation
in a particular Lorentz frame, i.e.,
in the rest frame for time-like $p$.
That is, only if the same value of $p\circ p$
as given by (\ref{eqn:3-13})
is substituted into the imaginary part, (\ref{eqn:3-8}),
the unitarity relation holds.
If different value of $p\circ p$ is substituted into the 
imaginary part, (\ref{eqn:3-8}),
the unitarity relation
is no longer valid.
In fact, the value of $p\circ p$ in the unitarity
sum would then be different from that in the imaginary
part, (\ref{eqn:3-8}),
which can be arbitrarily given.
This is an inconsistency of Gomis-Mehen's unitarity relation
that
twice the
imaginary part, (\ref{eqn:3-8})
is equal to
the unitarity sum, (\ref{eqn:3-9}),
for $p^2>4m^2$.
Such inconsistency never occurs if $p\circ p$
is Lorentz-invariant.
This is the case if $\theta^2_{\mu\nu}\equiv
\theta_{\mu\rho}\theta^{\rho}_{\;\;\nu}$
is a (symmetric) second-rank tensor,
or if $\theta^{\mu\nu}$ is an (antisymmetric) second-rank tensor.
\\
\ind
Let us now prove that $\theta^{\mu\nu}$ in the
$\theta$ algebra cannot be a nontrivial $c$-number tensor
provided $\hx^\mu$ is assumed to be a 4-vector.\footnote{This
transformation property is necessary to
define Lorentz-covariant field.}
To see this, 
let $U(\Lambda)$ be the unitary operator
of the Lorentz transformation
$$
\hx'{}^\mu=
U(\Lambda)\hx^\mu U^{-1}(\Lambda)
=\Lambda^\mu_{\;\;\nu}\hx^\nu.
$$
Sandwiching both sides of (\ref{eqn:2-1})
between the unitary operator $U(\Lambda)$ and its inverse,
we have the following for a $c$-number $\theta^{\mu\nu}$:
\beq
[\hx'{}^\mu,\hx'{}^\nu]=
\Lambda^\mu_{\;\;\rho}\Lambda^\nu_{\;\;\sigma}
[\hx^\rho,\hx^\sigma]
=\Lambda^\mu_{\;\;\rho}\Lambda^\nu_{\;\;\sigma}
i\theta^{\rho\sigma}=i\theta^{\mu\nu}.
\label{eqn:3-16}
\eeq
This equation holds only if $\theta^{\mu\nu}=0$
for $\Lambda^\mu_{\;\;\nu}=\delta^\mu_{\;\;\nu}+\omega^\mu_{\;\;\nu},
\;\;
\omega_{\mu\nu}=-\omega_{\nu\mu}$.
In fact,
(\ref{eqn:3-16}) for infinitesimal $\omega_{\mu\nu}$
can be cast into the form $
\omega_{\rho\sigma}f^{\rho\mu\sigma\nu}=0$,
which implies $
f^{\rho\mu\sigma\nu}
\equiv g^{\rho\mu}\theta^{\sigma\nu}
+g^{\rho\nu}\theta^{\mu\sigma}=0.$
Putting $\rho=\nu\ne \mu$ leads to $\theta^{\mu\sigma}=0$.
This only reflects the well-known fact that there
is no constant antisymmetric second-rank tensor.
\\
\ind
To summarize the imaginary part
of the amplitude (\ref{eqn:3-5}) in NC $\phi^3$ model
even if $\theta^{0i}=0$ \footnote{The integral
(\ref{eqn:3-10}) can also be done in a frame $\theta^{0i}=0$
with the same result (\ref{eqn:3-15})
provided $|\bk|=|\bk_{\rm cm}|$, where $p\circ p=(\theta^{ji}p_j)^2$
instead of (\ref{eqn:3-13}).}
cannot be obtained
by the usual prescription of
taking the UV limit $\Lambda^2\to \infty$ only.
To bypass this point
it is necessary to
adopt the Lorentz-covariant algebra
providing tensor nature of the
non-commutativity parameter
and to take into account
a consequent presence of the length parameter $a$
defined by (\ref{eqn:1-2}).
The latter is incorporated
in the new UV limit defined in (\ref{eqn:1-1}).
It means that
$p\circ p$ in (\ref{eqn:3-5})
should be consistently neglected in the limit
$\Lambda^2\to \infty$
because $p\circ p$ is of order $a^4$.
Then we only recover the commutative result
\bea*
{\rm Im}\,M&=&\frac{\lambda^2}{32\pi}\rho=\frac 12\sum|M|^2,
\eea*
where the right-hand side is calculated by taking the 
commutative limit in (\ref{eqn:3-9})
in accordance with (\ref{eqn:1-1}).
A single, Lorentz-scalar,
length parameter $a$
can be introduced into the theory
only if  $\theta^{\mu\nu}$
is assumed to be a tensor
so that $p\circ p$ is of order $a^4$.
This assumption is not tractable in the $\theta$
algebra as proved above.
To escape from this dilemma
we employ the DFR algebra (\ref{eqn:2-5})
and assume the Lorentz-invariant action (\ref{eqn:2-11}). 
\section{A new regularization based on Lorentz-invariant
NCQFT} 
It is now apparent that,
to avoid the problems issued in the previous section,
we have to clarify an invariant meaning
of the IR/UV mixing,
considering a Lorentz-invariant
version of the model.
\\
\ind
The Lorentz-invariant action of the NC $\phi^3$ 
model is given by
\beq
\hS&=&\int\!d^{\,6}\theta\, W(\theta)
\int\!d^{\,4}x[\frac 12\partial_\mu\phi(x)*
\partial^\mu\phi(x)-\frac 12m^2\phi(x)*\phi(x)
-\frac {\lambda}{3!}\phi(x)*
\phi(x)*\phi(x)],
\label{eqn:4-1}
\eeq
where the scalar field $\phi(x)$ is assumed\cite{4)} to be
`independent' of $\theta$. Using (\ref{eqn:1-2}) we put
\beq
W(\theta)&=&a^{-12}w({\bar\theta}).
\label{eqn:4-2}
\eeq
This converts (\ref{eqn:4-1}) into the form
\beq
\hS&=&\int\!d^{\,6}{\bar\theta}\, w({\bar\theta})
\int\!d^{\,4}x[\frac 12\partial_\mu\phi(x)*
\partial^\mu\phi(x)-\frac 12m^2\phi(x)*\phi(x)
-\frac {\lambda}{3!}\phi(x)*
\phi(x)*\phi(x)].
\label{eqn:4-3}
\eeq
The vertex in the Feynman diagram derived from 
the action (\ref{eqn:4-2}) is associated with
$-i\lambda$ times the vertex factor
\beq
V(k_1,k_2)&=&\int\!d^{\,6}{\bar\theta}\, w({\bar\theta})
\cos{(\frac{k_1\wedge k_2}2)}
\equiv \langle \cos{(\frac{k_1\wedge k_2}2)}\rangle,
\label{eqn:4-4}
\eeq
where $k_1$ and $k_2$ are the momenta flowing into
the vertex. Employing the action
(\ref{eqn:4-1}) simply means the replacement
of the vertex factor obtained by the action
(\ref{eqn:3-1}),
$\cos{(\frac{k_1\wedge k_2}2)}$, with
$\langle \cos{(\frac{k_1\wedge k_2}2)}\rangle$.
This makes it clear that
the identity $\cos^2{(\frac 12p\wedge q)}=
\frac{1+\cos{(p\wedge q)}}2$ responsible for
the one-loop amplitude (\ref{eqn:3-2})
to divide into the planar and nonplanar diagrams
can no longer be used in the Lorentz-invariant model.
\\
\ind
Using the above Feynman rule we obtain
the one-loop amplitude for the self-energy diagram
\beq
i\cM(p^2)
&=&\frac{\lambda^2}2\int\!\frac{d^{\,4}l}{(2\pi)^4}
\frac{V^2(p,l)}{((p-l)^2-m^2+i\epsilon)(l^2-m^2+i\epsilon)}.
\label{eqn:4-5}
\end{eqnarray}
There are some general properties of the amplitude
(\ref{eqn:4-5}). Because it is Lorentz-invariant,
it is a function of $p^2$ only provided
the integral is convergent. 
Let us compute (\ref{eqn:4-5}) only in a region where
the integral is finite. The precise meaning
of this assumption is made clear shortly.
Moreover, the integrand is a function of
$p^2, l^2$ and $p\cdot l$ by Lorentz invariance.
Using Feynman parameter, (\ref{eqn:4-5}) can be written as
\beq
\cM(p^2)
&=&-i\frac{\lambda^2}2\int\!\frac{d^{\,4}l}{(2\pi)^4}\int_0^1\!dx
\frac{V^2(p,l)}{(l^2-\Delta(p^2)+i\epsilon)^2},
\label{eqn:4-6}
\end{eqnarray}
where the translation of the integration variable $l$
is made.\
Next step consists of Wick rotation,
\beq
l^0=il_E^4,&&\qquad
\bl=\bl_E,\nn\\[2mm]
p^0=ip_E^4,&&\qquad
\bp=\bp_E,
\label{eqn:4-7}
\end{eqnarray}
where Euclidean momenta are real.\footnote{The
Wick rotation with respect to $l$
is allowed if the contributions
from the large arcs in
$1^{\rm st}$ and 
$3^{\rm rd}$ quadrants in
the $l_0$-plane
can be neglected.
This is assured if the integration
over $\theta$ is performed after that over $l$
in a frame,
$p_0\ne 0, \bp={\mbf 0}$, so that
$\langle\cos{(\frac 12p\wedge l)}\rangle=
\langle\cos{(\frac 12p_0\theta^{0i}l_i)}\rangle$
does not contain $l_0$.
Next perform the Wick rotation with respect to $p$ 
after covariantizing the result.}
The result
turns out to be
\beq
\cM(-p_E^2)
&=&\frac{\lambda^2}2\int\!\frac{d^{\,4}l_E}{(2\pi)^4}\int_0^1\!dx
\frac{V^2(p_E,l_E)}{[l_E^2+\Delta(-p_E^2)]^2}.
\label{eqn:4-8}
\end{eqnarray}
Note that $(p_E\cdot l_E)^2$
is now equal to $p_E^2l_E^2\cos^2{\theta_1}$
where $\theta_1$ is the angle between the two vectors,
$p_E$ and $l_E$.
The commutative limit
corresponds to putting
$V^2(p_E,l_E)=1$, that is, $p_E^2=0$,\footnote{Replace $k^2p^2
-(k\cdot p)^2$ in (\ref{eqn:4-16}) of Ref.~6)
with $p^2l^2-(p\cdot l)^2$
and go over to Euclidean metric.}
in which case
the amplitude 
(\ref{eqn:4-8}) shows log divergence.
Thus the amplitude
(\ref{eqn:4-8}) must diverge at $p_E^2=0$,
i.e., in the IR limit.
We call it IR divergence.
In other words, 
finite amplitude would be obtained as far as $p_E^2\ne 0$
in Euclidean metric.
The conventional UV divergence
is translated to the IR divergence.
The physical reason is that
the IR limit cannot be distinguished
from the commutative limit
and Lorentz-invariant NCQFT satisfies the corresponding principle
so that it smoothly reduces to QFT in the
commutative limit with the well-known UV divergence.
\\
\ind
In order to evaluate the integral (\ref{eqn:4-8}) 
we must determine the vertex factor $V(p_E,l_E)$.
Because the weight function is even, $w(-\theta)=w(\theta)$,
by Lorentz invariance, (\ref{eqn:4-4})
may be written as
\beq
V(k_1,k_2)&=&\langle e^{\frac i2{k_1\wedge k_2}}\rangle.
\label{eqn:4-9}
\eeq
The average $\langle\cdots\rangle$ 
can be calculated
once the weight function is given.
\\
\ind
There is no guiding principle to determine the weight function.
Nonetheless, the presence of the IR
divergence does not depend on the precise form
of it as argued above. This in turn allows us to adopt
a most convenient form.
In the following calculation the non-commutativity parameter
is also made Euclidean,
\beq
\theta^{0i}\to -i\theta_E^{4i},\;\;\;
\theta^{ij}\to \theta_E^{ij}
\label{eqn:4-10}
\eeq
which corresponds to positive 
${\bar\alpha}=\frac 12{\bar\theta^{\mu\nu}}
{\bar\theta_{\mu\nu}}$,
such that
\beq
p\wedge l=p_E\wedge_El_E\equiv
\sum_{\mu,\nu=1,2,3,4}(p_E)_\mu\theta_E^{\mu\nu}(l_E)_\nu.
\label{eqn:4-11}
\eeq
Assuming the Gaussian weight function\cite{6)}
\begin{eqnarray}
w({\bar\theta}_E)=\frac 1{\pi^3}
e^{-b[({\bar\theta}^{41})^2+({\bar\theta}^{42})^2+({\bar\theta}^{43})^2
+({\bar\theta}^{12})^2
+{\bar\theta}^{23})^2+({\bar\theta}^{31})^2]}, b>0,
\label{eqn:4-12}
\end{eqnarray}
the vertex factor is determined as
\beq
V(p_E,l_E)&=&
e^{-\frac{A_E}2[l_E^2p_E^2-(p_E\cdot l_E)^2]},
\label{eqn:4-13}
\end{eqnarray}
where
\beq
A_E=\frac {a^4}2\frac{\langle {\bar\theta}^{\,2}_E\rangle}{24}
\label{eqn:4-14}
\eeq
with $\langle {\bar\theta}^{\,2}_E\rangle=6/b$.
The vertex factor (\ref{eqn:4-13})
is called the invariant
damping factor in Ref.~6).
Substituting (\ref{eqn:4-13}) into
(\ref{eqn:4-8}) and using Schwinger representation
yield
\beq
\cM(-p_E^2)
&=&\frac{\lambda^2}{32\pi^4}
\int_0^1\!dx\int_0^\infty\! dss
\int\!d^{\,4}l_E
e^{-s(l_E^2+\Delta(-p_E^2))-A_E[l_E^2p_E^2-(p_E\cdot l_E)^2]}.
\label{eqn:4-15}
\end{eqnarray}
The $l_E$-integration is easily done
choosing the direction of the vector $p_E$ as the polar axis in
$l_E$-space with the result
\beq
\cM(-p_E^2)
&=&\frac{\lambda^2}{32\pi^2}
\int_0^1\!dx\int_0^\infty\! ds
\frac{\sqrt{s}}{\big(\sqrt{s+A_Ep_E^2}\big)^3}
e^{-s\Delta(-p_E^2)}.
\label{eqn:4-16}
\eeq
As promised this integral is finite
outside the IR limit $p_E^2=0$.
However,
it is divergent at $p_E^2=0$.
It is crucial that
there is no distinction
between the planar and non-planar
diagrams in the Lorentz-invariant NCQFT, 
making it feasible to
`subtract off' the IR singularity
from the total amplitude in relation
to the subtraction of the UV divergence.
The commutative
limit $a^2\to 0$
recovers the well-known
UV divergent amplitude
obtained by putting $p\circ p=0$
in (\ref{eqn:3-3})
without the regularization factor.\footnote{The regularization factor
in the present model is introduced to avoid the IR singularity, see below.}
\\
\ind
The IR singularity in (\ref{eqn:4-16})
is a branch point as in the previous model, (\ref{eqn:3-6}).
To see this we note that
the $s$-integral in (\ref{eqn:4-16}) is expressed in terms of
the Whittaker function $W_{\lambda,\mu}(z)$,
\bea*
\int_0^\infty\! ds
\frac{\sqrt{s}}{\big(\sqrt{s+A_Ep_E^2}\big)^3}
e^{-s\Delta(-p_E^2)}
&=&
[A_Ep_E^2\Delta(-p_E^2)]^{-\frac 12}\Gamma(\frac 32)
W_{-1,0}\big(A_Ep_E^2\Delta(-p_E^2)\big).
\eea*
Using the expansion\cite{14)} of the Whittaker function
the $s$-integral turns out to be
\bea*
\int_0^\infty\! ds
\frac{\sqrt{s}}{\big(\sqrt{s+A_Ep_E^2}\big)^3}
e^{-s\Delta(-p_E^2)}
&=&
\frac 1{\Gamma(\frac 32)}
\sum_{k=0}^\infty
\frac{\Gamma(k+\frac 32)}
{(k!)^2}
[A_Ep_E^2\Delta(-p_E^2)]^k\nn\\[2mm]
&&\times
\{2\psi(k+1)-\psi(k+\frac 32)-\ln{[A_Ep_E^2\Delta(-p_E^2)]}\},
\eea*
where $\psi(z)=\Gamma'(z)/\Gamma(z)$.
This equation indicates that
the IR singularity at $p_E^2=0$
in (\ref{eqn:4-16}) is a branch point.\footnote{The branch
point singularity at $\Delta(p^2)=0$
is associated with the unitarity.}
The above expression is not convenient
for the subtraction of the IR singularity
because the IR limit cannot
be distinguished from the commutative limit
in (\ref{eqn:4-16}),
while the equation
$\ln{(A_Ep_E^2)}=\ln{A_E}+\ln{p_E^2}$
obscures this simple fact.
Thus we do not use the Whittaker function in this paper.
\\
\ind
The amplitude
(\ref{eqn:4-16}) goes over in Minkowski space
to
\beq
\cM_{\Lambda^2}(p^2)
&=&\frac{\lambda^2}{32\pi^2}
\int_0^1\!dx\int_{1/\Lambda^2}^\infty\! ds
\frac{\sqrt{s}}{\big(\sqrt{s-Ap^2}\big)^3}
e^{-s\Delta(p^2,m^2)},
\label{eqn:4-17}
\eeq
where
\beq
A=\frac {a^4}2\frac{\langle {\bar\theta}^{\,2}\rangle}{24}
\label{eqn:4-18}
\eeq
is positive (remember ${\bar\alpha}>0$)
so that
the singularity at $s=Ap^2$ for time-like $p$
should be expelled outside
the integration region over $s$.
To accomplish this the lower limit of the integration
over $s$
is put to $1/\Lambda^2$ where
$\Lambda$ is the UV cutoff of order $a^{-1}$.
Because $A$ is of order $a^4$,
$Ap^2$ in the integrand
can be neglected in the new UV limit, leaving the integral
\beq
\cM_{\Lambda^2}(p^2)
&=&\frac{\lambda^2}{32\pi^2}
\int_0^1\!dx\int_{1/\Lambda^2}^\infty\! dss^{-1}
e^{-s\Delta(p^2,m^2)}
=-\frac{\lambda^2}{32\pi^2}\int_0^1\!dx
{\rm Ei}(-\Delta(p^2,m^2)/\Lambda^2),
\label{eqn:4-19}
\eeq
where Ei$(z)$ is the exponential function.
(Recall that $\Delta(p^2,m^2)>0$ is assumed.)
Using the relation Ei$(-z)\to \ln{z}+\gamma$
as $z\to 0$,
the amplitude becomes in the new UV limit
\beq
\cM_{\Lambda^2}(p^2)
&\to&-\frac{\lambda^2}{32\pi^2}\int_0^1\!dx
\big(\gamma+\ln{(\Delta(p^2,m^2)/\Lambda^2)}\big).
\label{eqn:4-20}
\eeq
The log divergence\footnote{The divergent behavior
is the same as seen from (\ref{eqn:3-5})
with $p\circ p=0$ except for $2\gamma\to \gamma$
in the latter.} must be subtracted off to
define the well-defined amplitude (mass renormalization).
For instance, we define the renormalized amplitude 
by subtraction at $p^2=\mu^2$:
\beq
\cM_R(p^2,\mu^2)
&=&\mathop{\rm lim}_{\Lambda^2\to\infty,
a^2\to 0, \Lambda^2a^2:{\rm fixed}}\;
[\cM_{\Lambda^2}(p^2)-\cM_{\Lambda^2}(\mu^2)]\nn\\[2mm]
&=&-\frac{\lambda^2}{32\pi^2}
\int_0^1\!dx
\ln{\frac {\Delta(p^2,m^2)}{\Delta(\mu^2,m^2)}}.
\label{eqn:4-21}
\eeq
\ind
In passing we remark that the same result
is obtained by defining
\beq
\cM_{\Lambda^2}(p^2)
&=&\frac{\lambda^2}{32\pi^2}
\int_0^\infty\!dx\int_0^\infty\! ds
\frac{\sqrt{s}}{\big(\sqrt{s-Ap^2}\big)^3}
e^{-s\Delta(p^2,m^2)-\frac 1{s\Lambda^2}},
\label{eqn:4-22}
\eeq
for space-like $p$
with the expansion
\beq
\frac{\sqrt{s}}{\big(\sqrt{s-Ap^2}\big)^3}
=\sum_{n=0}^\infty
\left(
\ba{c}
-\frac 32\\
n\\
\ea
\right)
s^{-1-n}(-Ap^2)^n
\label{eqn:4-23}
\eeq
inside the integral.
The $s$-integral
in the $n$-th term
of the expansion
is given by the modified Bessel
function, $2(\sqrt{\Delta\Lambda^2})^n
K_n\big(2\sqrt{\frac \Delta{\Lambda^2}}\big)$,
where $\Delta=\Delta(p^2,m^2)$.
The behavior of the function
$K_n(z)$ at $z\to 0$
shows that
only the first term in the expansion (\ref{eqn:4-23})
survives in the new UV limit
in accordance with (\ref{eqn:4-20})
except for $2\gamma$ instead of $\gamma$.
Analytic continuation to positive $p^2$
can be done after the new UV limit.
Euler's constant does not appear
in the subtracted form.
Consequently, in what follows we employ the second definition
(\ref{eqn:4-22}).
\section{Vacuum polarization in Lorentz-invariant NCQED} 
The purpose of this section
is to prove the gauge invariance of the regularization
in the previous section by computing
the vacuum polarization in Lorentz-invariant NCQED.\cite{4)}
\\
\ind
The Lorentz-invariant NCQED is defined by the action
\beq
\hS_{\rm D}
&=&
\int\!d^4xd^6\theta w(\btheta)\big[{\bar\psi}(x)
(i\gamma^\mu\partial_\mu-M)\psi(x)+
e{\bar\psi}(x)*
\gamma^\mu A_\mu(x)*\psi(x)\big].
\label{eqn:5-1}
\eeq
We use the action (\ref{eqn:2-11}),
assuming that the spinor and the gauge field
are `independent' of $\theta$.
The relevant gauge transformations are:
\beq
\psi(x)&\to& ^{\hat g}\psi(x)
=U(x)*\psi(x),\nn\\[2mm]
{\bar\psi}(x)&\to& ^{\hat g}{\bar\psi}(x)
={\bar\psi}(x)*U^{\dag}(x),
\label{eqn:5-2}
\eeq
where $U(x)$ is assumed to be $*$-unitary:
\beq
U(x)*U^{\dag}(x)=
U^{\dag}(x)*U(x)=1.
\label{eqn:5-3}
\eeq
The $*$-gauge invariance is proved by
the transformation property of the NC gauge field,
\beq
A_\mu(x)\to
^{\hat g}\!\!A_\mu(x)&=&U(x)*A_\mu(x)*
U^{\dag}(x)+\frac ieU(x)*\partial_\mu U^{\dag}(x).
\label{eqn:5-4}
\eeq
The Maxwell sector will be considered in the next section.
\\
\ind
Using the action (\ref{eqn:5-1}) the vacuum
polarization tensor
is given by
\beq
i\Pi^{\mu\nu}(q)&=&
(ie)^2(-1)\int\!\frac{d^4l}{(2\pi)^4}
{\rm Tr}\big[
\gamma^\mu\frac i{\lslash-M+i\epsilon}
\gamma^\nu\frac i{\qslash+\lslash-M+i\epsilon}\big]
\langle e^{\frac i2q\wedge l}\rangle
\langle e^{\frac i2l\wedge q}\rangle,
\label{eqn:5-5}
\eeq
where $q$ is the external photon momentum.
The case of the Lorentz-non-invariant NCQED
is obtained by replacing the average
$\langle e^{\frac i2q\wedge l}\rangle$
with $e^{\frac i2q\wedge l}$.
Without average brackets the last two factors of
the right-hand side in
(\ref{eqn:5-5}) cancel out and the result is the same\cite{7)}
as in the ordinary QED.
Such cancellation does not occur
in the Lorentz-invariant NCQED,
leading to a highly non-trivial result.
\\
\ind
By multiplying $q_\mu$ with the tensor (\ref{eqn:5-5})
gives
\beq
q_\mu\Pi^{\mu\nu}(q)&=&
ie^2\int\!\frac{d^4l}{(2\pi)^4}
{\rm Tr}\big[
\qslash\frac 1{\lslash-M+i\epsilon}
\gamma^\nu\frac 1{\qslash+\lslash-M+i\epsilon}\big]V^2(q,l)\nn\\[2mm]
&=&ie^2\int\!\frac{d^4l}{(2\pi)^4}
{\rm Tr}\big[\frac 1{\lslash-M+i\epsilon}
\gamma^\nu-\gamma^\nu\frac 1{\qslash+\lslash-M+i\epsilon}\big]V^2(q,l).
\label{eqn:5-6}
\eeq
Because $V^2(q,l)$ acts as a damping factor
(at least in Euclidean metric)
as in the scalar model in the previous section,
it is possible to translate the integration variable
in the second integral, 
$l\to l-q$ to prove the transversality
\beq
q_\mu\Pi^{\mu\nu}(q)&=&0.
\label{eqn:5-7}
\eeq
The gauge invariance (\ref{eqn:5-7})
is explicitly proved as follows.
\\
\ind
As usual, we compute the Dirac trace,
\beq
{\rm Tr}\big[
\gamma^\mu\frac i{\lslash-M+i\epsilon}
\gamma^\nu\frac i{\qslash+\lslash-M+i\epsilon}\big]
=\frac{-N^{\mu\nu}}{(l^2-M^2+i\epsilon)
((q+l)^2-M^2+i\epsilon)},
\label{eqn:5-8}
\eeq
where
\beq
N^{\mu\nu}&=&4[l^\mu(q+l)^\nu-
g^{\mu\nu}l\cdot(q+l)+l^\nu(q+l)^\mu
+g^{\mu\nu}M^2].
\label{eqn:5-9}
\eeq
Combining the two denominators
by the Feynman parameter
(\ref{eqn:5-8}) becomes
\beq
-\int_0^1\!dx\frac{N^{\mu\nu}}
{[(l+q(1-x))^2-\Delta(q^2,M^2)+i\epsilon]^2}.
\label{eqn:5-10}
\eeq
Translation of the integration variable,
$l\to l-q(1-x)$, brings the vacuum polarization tensor to the form
\beq
\Pi^{\mu\nu}(q)&=&
ie^2\int_0^1\!dx\int\!\frac {d^4l}
{(2\pi)^4}
\frac{{N'}{}^{\mu\nu}}{[l^2-\Delta(q^2,M^2)+i\epsilon]^2}
V^2(q,l),
\label{eqn:5-11}
\eeq
where we have used the relation
$V^2(q,l-q(1-x))=V^2(q,l)$ and $N^{\mu\nu}$ equals
\beq
{N'}{}^{\mu\nu}&=&
4[2l^\mu l^\nu-
g^{\mu\nu}(l^2-\Delta(q^2,M^2))
+(g^{\mu\nu}q^2-q^\mu q^\nu)2x(1-x)]
\label{eqn:5-12}
\eeq
up to terms linear in $l^\mu$.
The linear terms drop out because
$V^2(q,l)$ is even in $l$.
The Wick rotation is performed at this stage:
\beq
l^0&=&il_E^4, \quad {\bl}={\bl}_E\nn\\[2mm]
q^0&=&iq_E^4, \quad {\bq}={\bq}_E.
\label{eqn:5-13}
\eeq
The vacuum polarization tensor becomes in
Euclidean metric
\beq
\Pi^{\mu\nu}_E(q_E)&=&
-e^2\int_0^1\!dx\int\!\frac {d^4l_E}
{(2\pi)^4}
\frac{{N'}{}_E^{\mu\nu}}{[l_E^2+\Delta(-q_E^2,M^2)]^2}
V^2(q_E,l_E),
\label{eqn:5-14}
\eeq
where
\beq
{N'}{}_E^{\mu\nu}=4[2l_E^\mu l_E^\nu+
g_E^{\mu\nu}(l_E^2+\Delta(-q_E^2,M^2))
+(-g_E^{\mu\nu}q_E^2-q_E^\mu q_E^\nu)2x(1-x)].
\label{eqn:5-15}
\eeq
\ind
Omitting the index $E$ for typographical reason
till (\ref{eqn:5-29}),
we separate the amplitude (\ref{eqn:5-14})
into two parts,
\beq
\Pi^{\mu\nu}&=&
\stackrel{\!\!\!(1)}{\Pi^{\mu\nu}}+
\stackrel{\!\!\!(2)}{\Pi^{\mu\nu}},\nn\\[2mm]
\stackrel{\!\!\!(1)}{\Pi^{\mu\nu}}&=&
-4e^2\int_0^1\!dx\int\!\frac{d^4l}{(2\pi)^4}
\frac{2l^\mu l^\nu+g^{\mu\nu}(l^2+\Delta)}{(l^2+\Delta)^2}
V^2(q,l),\nn\\[2mm]
\stackrel{\!\!\!(2)}{\Pi^{\mu\nu}}&=&
(-g^{\mu\nu}q^2-q^\mu q^\nu)\stackrel{(2)}{\Pi}(-q^2),\nn\\[2mm]
\stackrel{(2)}{\Pi}(-q^2)&=&
-4e^2\int_0^1\!dx2x(1-x)\int\!\frac{d^4l}{(2\pi)^4}
\frac{V^2(q,l)}
{(l^2+\Delta)^2},
\label{eqn:5-16}
\eeq
with $\Delta\equiv\Delta(-q^2,M^2)$.
We now show that
$\stackrel{\!\!\!(1)}{\Pi^{\mu\nu}}$
is also transverse, that is,
it is proportional to the tensor,
$(-g^{\mu\nu}q^2-q^\mu q^\nu)$.
The reason that the term proportional to
$q^\mu q^\nu$,
which is absent in QED and Lorentz-non-invariant
NCQED,
appears in the Lorentz-invariant NCQED
lies in the fact that
the vertex factor $V(q,l)$ is a function of
not only $l^2, q^2$ but also $l\cdot q$.
Putting
\beq
\int\!\frac{d^4l}{(2\pi)^4}
\frac{l^\mu l^\nu}{(l^2+\Delta)^2}
V^2(q,l)&=&
C_1g^{\mu\nu}+C_2q^\mu q^\nu
\label{eqn:5-17}
\eeq
and
\beq
\int\!\frac{d^4l}{(2\pi)^4}
\frac{V^2(q,l)}{l^2+\Delta}
&=&C_3,
\label{eqn:5-18}
\eeq
where $C_{1,2,3}$ are functions of the invariant, $q^2$,
we obtain
\beq
\stackrel{\!\!\!(1)}{\Pi^{\mu\nu}}
&=&-4e^2\int_0^1\!dx\big[(2C_1+C_3)g^{\mu\nu}+2C_2q^{\mu\nu}\big]
\nn\\[2mm]
&=&(-g^{\mu\nu}q^2-q^\mu q^\nu)\stackrel{(1)}{\Pi}(-q^2)
\nn\\[2mm]
\stackrel{(1)}{\Pi}(-q^2)
&=&4e^2\int_0^1\!dx(2C_2),
\label{eqn:5-19}
\eeq
provided that
\beq
2C_1+C_3=2C_2q^2.
\label{eqn:5-20}
\eeq
The proof of this equation goes through as follows.
To integrate (\ref{eqn:5-17}) choose the 4-th direction
in $l$-space as pointing to the vector
$q$ so that
\beq
q&=&(0,0,0,q),\quad l=(l^1,l^2,l^3,l^4),\nn\\[2mm]
l^4&=&l\cos{\theta_1},\quad l^3=l\sin{\theta_1}\cos{\theta_2},\nn\\[2mm]
l^2&=&l\sin{\theta_1}\sin{\theta_2}\cos{\theta_3},
\quad l^1=l\sin{\theta_1}\sin{\theta_2}\sin{\theta_3}.
\label{eqn:5-21}
\eeq
The $\mu=\nu=4$ component of (\ref{eqn:5-17})
is then given by
\beq
\int\!\frac{d^4l}{(2\pi)^4}
\frac{l^2\cos^2{\theta_1}}{(l^2+\Delta)^2}
V^2(q,l)&=&
-C_1+C_2q^2,
\label{eqn:5-22}
\eeq
because $g^{44}=-1$.
Noting (\ref{eqn:4-13}) which reads in the present notation
\beq
V(q,l)&=&
e^{-\frac{A}2[q^2l^2-(q\cdot l)^2]}
=e^{-\frac{A}2q^2l^2(1-\cos^2{\theta_1})}
=e^{-\frac{A}2q^2l^2\sin^2{\theta_1}},
\label{eqn:5-23}
\end{eqnarray}
the left-hand side of (\ref{eqn:5-22})
becomes
\beq
\int\!\frac{d^4l}{(2\pi)^4}
\frac{l^2\cos^2{\theta_1}}{(l^2+\Delta)^2}
e^{-Aq^2l^2\sin^2{\theta_1}}&=&
\frac 1{4\pi^3}
\int_0^\infty\!ds se^{-s\Delta}
\int_0^\pi\!d\theta_1\cos^2{\theta_1}
\sin^2{\theta_1}
\int_0^\infty\!dll^5e^{-(s+Aq^2\sin^2{\theta_1})l^2}
\nn\\[2mm]
&=&\frac 1{4\pi^3}
\int_0^\infty\!ds se^{-s\Delta}
\int_0^\pi\!d\theta_1
\frac {\cos^2{\theta_1}
\sin^2{\theta_1}}{[s+Aq^2\sin^2{\theta_1}]^3}\nn\\[2mm]
&=&
\frac 1{32\pi^2}
\int_0^\infty\!ds 
\frac {e^{-s\Delta}}{\sqrt{s}\big(\sqrt{s+Aq^2}\big)^3}.
\label{eqn:5-24}
\end{eqnarray}
Namely, we have
\beq
-C_1+C_2q^2=\frac 1{32\pi^2}
\int_0^\infty\!ds 
\frac {e^{-s\Delta}}
{\sqrt{s}\big(\sqrt{s+Aq^2}\big)^3}.
\label{eqn:5-25}
\end{eqnarray}
The $\mu=\nu=3$ component of (\ref{eqn:5-17})
using (\ref{eqn:5-21}) is given by
\beq
\int\!\frac{d^4l}{(2\pi)^4}
\frac{l^2\cos^2{\theta_1}\sin^2{\theta_1}}{(l^2+\Delta)^2}
e^{-Aq^2l^2\sin^2{\theta_1}}&=&
-C_1,
\label{eqn:5-26}
\eeq
because $g^{33}=-1$. Calculating the left-hand side
as in (\ref{eqn:5-24})
we find that
\beq
C_1=-\frac 1{32\pi^2}
\int_0^\infty\!ds 
\frac {\sqrt{s}e^{-s\Delta}}{\big(\sqrt{s+Aq^2}\big)^5}.
\label{eqn:5-27}
\end{eqnarray}
Adding the two equations (\ref{eqn:5-25})
and (\ref{eqn:5-27}) $C_2$ is determined to be
\beq
C_2&=&
\frac 1{32\pi^2q^2}
\int_0^\infty\!ds 
\frac {e^{-s\Delta}}{\sqrt{s}}
\big[
\frac 1{\big(\sqrt{s+Aq^2}\big)^3}
-\frac s{\big(\sqrt{s+Aq^2}\big)^5}\big]\nn\\[2mm]
&=&
\frac 1{32\pi^2}A
\int_0^\infty\!ds 
\frac {e^{-s\Delta}}
{\sqrt{s}\big(\sqrt{s+Aq^2}\big)^5}.
\label{eqn:5-28}
\end{eqnarray}
On the other hand, $C_3$ is given by
\beq
C_3&=&
\int\!\frac{d^4l}{(2\pi)^4}
\frac 1{l^2+\Delta}
e^{-Aq^2l^2\sin^2{\theta_1}}\nn\\[2mm]
&=&\frac 1{16\pi^2}
\int_0^\infty\!ds
\frac{e^{-s\Delta}}{\sqrt{s}\big(\sqrt{s+Aq^2}\big)^3}.
\label{eqn:5-29}
\end{eqnarray}
Comparing this equation with
(\ref{eqn:5-25})
leads to (\ref{eqn:5-20})
which is to be proved.\footnote{The proof hinges upon
the Gaussian weight function, but the transversality
(\ref{eqn:5-7}) was proved for general weight function.
Hence we conjecture that (\ref{eqn:5-20})
holds true in general.}
\\
\ind
Analytic continuation back to Minkowski space gives
(from here on we use the Lorentz metric)
\beq
\stackrel{\!\!\!({i})}{\Pi^{\mu\nu}}&=&
(g^{\mu\nu}q^2-q^\mu q^\nu)\stackrel{({i})}
{\Pi}_{\Lambda^2}(q^2),\quad i=1,2,\nn\\[2mm]
\stackrel{(1)}{\Pi}_{\Lambda^2}(q^2)&=&\frac \alpha\pi A
\int_0^1\!dx\int_0^\infty\!ds 
\frac {e^{-s\Delta(q^2,M^2)-\frac 1{s\Lambda^2}}}
{\sqrt{s}(\sqrt{s-Aq^2})^5},
\nn\\[2mm]
\stackrel{(2)}{\Pi}_{\Lambda^2}(q^2)&=&
-\frac \alpha\pi\int_0^1\!dx2x(1-x)
\int_0^\infty\!ds\frac{\sqrt{s}e^{-s\Delta(q^2,M^2)
-\frac 1{s\Lambda^2}}}
{(\sqrt{s-Aq^2})^3},
\label{eqn:5-30}
\end{eqnarray}
where $\alpha=e^2/4\pi$ and
we have inserted the regularization factor
$e^{-\frac 1{s\Lambda^2}}$
to remove the IR divergence
by subtraction in the next stage.
Expanding the integrands as in (\ref{eqn:4-23})
and performing $s$-integrals
give in the new UV limit (\ref{eqn:1-1})
after subtraction at $q^2=0$ (charge renormalization),\footnote{It can
be shown that $\stackrel{({1})}
{\Pi}_{\Lambda^2}(q^2)\to 0$ in the new UV limit
if $\Lambda^2a^2\to 0$.}
\beq
\stackrel{(1)}{\Pi}_{\lower 1.2pt \hbox{{\scriptsize ${R}$}}}(q^2)
&=&\mathop{\rm lim}_{\Lambda^2\to\infty,
a^2\to 0, \Lambda^2a^2:{\rm fixed}}\;
[\stackrel{(1)}{\Pi}_{\Lambda^2}(q^2)-
\stackrel{(1)}{\Pi}_{\Lambda^2}(0)]=0,\nn\\[2mm]
\stackrel{(2)}{\Pi}_{\lower 1.2pt \hbox{{\scriptsize ${R}$}}}(q^2)
&=&\mathop{\rm lim}_{\Lambda^2\to\infty,
a^2\to 0, \Lambda^2a^2:{\rm fixed}}\;
[\stackrel{(2)}{\Pi}_{\Lambda^2}(q^2)-
\stackrel{(2)}{\Pi}_{\Lambda^2}(0)]\nn\\[2mm]
&=&-\frac{2\alpha}\pi
\int_0^1\!dxx(1-x)\ln{\frac{M^2}{\Delta(q^2,M^2)}}.
\label{eqn:5-31}
\eeq
This is the same result as obtained by
Pauli-Villars or dimensional regularizations.
It is a novel feature
of the Lorentz-invariant NCQED that
the non-transverse part of the vacuum
polarization tensor in QED,
which is to be shown to vanish in any gauge-invariant
regularization methods, is automatically transverse
without regularization.
Namely, in the Lorentz-invariant NCQED
neither unphysical fields nor analytic continuation
to complex dimension need be introduced
to calculate Feynman integral.
The extra dependence
of the vertex factor $V$ on the inner
product of two momenta incoming to the vertex
complicates, however,
computation of the electron vertex function and the triangle
diagram which accompany three $V$'s.
On the contrary, the electron self-energy
diagram is easily computed, because it
contains only two $V$'s.
\section{Tadpole diagrams and fermion loop
with Lorentz-invariant NC scalar coupling} 
In this section we continue to study the Lorentz-invariant
NCQED in the Maxwell sector. The relevant
action using (\ref{eqn:2-11}) is given by
\begin{eqnarray} 
{\hat S}_M
&=&-\displaystyle{{1\over 4}}\int\!d^4xd^6\btheta w(\btheta)
F_{\mu\nu}(x)*F^{\mu\nu}(x),
\label{eqn:6-1}
\end{eqnarray}
where the field strength
tensor is defined by
\begin{eqnarray} 
F_{\mu\nu}(x)&=&\partial_\mu A_\nu(x)
-\partial_\nu A_\mu(x)
      -ie[A_\mu(x),A_\nu(x)]_*,
\label{eqn:6-2}
\end{eqnarray}
with the Moyal bracket
\beq
[A_\mu(x),A_\nu(x)]_*\equiv
A_\mu(x)*A_\nu(x)-A_\nu(x)*A_\mu(x).
\label{eqn:6-3}
\end{eqnarray}
Here, the $*$-gauge transformation property
of the NC gauge field is given by (\ref{eqn:5-4}).
\\
\ind
In order to consistently quantize the gauge field
in NCQED
it is necessary to introduce the ghost fields, $c,{\bar c}$, and
the Nakanishi-Lautrup field $B$ such that the full action is
BRST-invariant.\cite{15)}
We use the Feynman rules given in Ref.~7)
and choose the Feynman-'t Hooft gauge.
Note, however, that there exist no three-point vertices
in the Lorentz-invariant NCQED if the action
(\ref{eqn:6-1}) is employed,
because $\langle\sin{(\frac 12p\wedge q)}\rangle=0$.
Consequently, there is only
one more contribution
to the photon self energy,
the tadpole diagram.
This greatly simplifies the computation.
\\
\ind
The tadpole diagram is
given by
\beq
i\Pi_{\rm tadpole}^{\mu\nu}(q^2)
&=&-24e^2g^{\mu\nu}
\int\!\frac{d^4l}{(2\pi)^4}
\frac 1{l^2+i\epsilon}\langle\sin^2{(\frac 12q\wedge l)}\rangle\nn\\[2mm]
&=&-12e^2g^{\mu\nu}
\big[\int\!\frac{d^4l}{(2\pi)^4}
\frac 1{l^2+i\epsilon}-\int\!\frac{d^4l}{(2\pi)^4}
\frac 1{l^2+i\epsilon}e^{-2A(q^2l^2-(q\cdot l)^2)}\big],
\label{eqn:6-4}
\eeq
where $q$ denotes the external photon momentum,
we have used the fact
$\langle 1\rangle=1$ and
\beq
\langle\cos{(q\wedge l)}\rangle
=V(\sqrt{2}q,\sqrt{2}l)=
e^{-2A((q^2l^2-(q\cdot l)^2)}.
\label{eqn:6-5}
\eeq
By Wick rotation the second integral in (\ref{eqn:6-4})
is given by
\beq
\int\!\frac{d^4l}{(2\pi)^4}
\frac 1{l^2+i\epsilon}e^{-2A(q^2l^2-(q\cdot l)^2)}
&=&
-i\int\!\frac{d^4l_E}{(2\pi)^4}
\frac 1{l_E^2}e^{-2A_E(q_E^2l_E^2-(q_E\cdot l_E)^2)}\nn\\[2mm]
&=&
-\frac i{16\pi^2}\int_0^\infty\!ds
\frac {e^{-s\lambda^2}}
{\sqrt{s}\big(\sqrt{s+2A_Eq_E^2}\big)^3},
\label{eqn:6-6}
\eeq
where $\lambda$ is a small photon mass.\footnote{This
is introduced here only for computational purpose.}
The integral (\ref{eqn:6-6})
shows the IR divergence
as before
and we introduce the UV cutoff to define
the regularized amplitude after going back to
Minkowski space
\beq
\Pi_{\rm tadpole, \Lambda^2}^{\mu\nu}(q^2)
&=&-12e^2g^{\mu\nu}
\big[\int\!\frac{d^4l}{(2\pi)^4}
\frac 1{i(l^2+i\epsilon)}|_{\Lambda^2}
+\frac 1{16\pi^2}\int_0^\infty\!ds
\frac {e^{-s\lambda^2-\frac 1{s\Lambda^2}}}
{\sqrt{s}\big(\sqrt{s-2Aq^2}\big)^3}\big],
\label{eqn:6-7}
\eeq
where
$|_{\Lambda^2}$
in the first integral
means a regularized integral
to be taken,
which, in fact,
is cancelled out by the second one
with $q^2=0$.
Thus we have
\beq
\Pi_{{\rm tadpole}, R}^{\mu\nu}(q^2)
&=&\mathop{\rm lim}_{\Lambda^2\to\infty,
a^2\to 0, \Lambda^2a^2:{\rm fixed}}\;
\Pi_{\rm tadpole, \Lambda^2}^{\mu\nu}(q^2)
\nn\\[2mm]
&=&
\mathop{\rm lim}_{\Lambda^2\to\infty,
a^2\to 0, \Lambda^2a^2:{\rm fixed}}\;
\big(-\frac{3\alpha}\pi\big)g^{\mu\nu}\nn\\[2mm]
&&\times
\int_0^\infty\!ds
\sum_{n=1}^\infty
e^{-s\lambda^2-\frac 1{s\Lambda^2}}
\left(
\ba{c}
-\frac 32\\
n\\
\ea
\right)
s^{-2-n}(2Aq^2)^n.
\label{eqn:6-8}
\eeq
In the new UV limit
only the first $n=1$ term
survives, whose $s$-integral is given by
\beq
\int\!dss^{-3}e^{-s\lambda^2-\frac 1{s\Lambda^2}}
=2(\lambda^2\Lambda^2)K_2(2\sqrt{\lambda^2/\Lambda^2})
\to
\Lambda^4,
\label{eqn:6-9}
\eeq
in the limit $\Lambda^2\to \infty$.
A finite result is then left over
in the new UV limit:
\beq
\Pi_{{\rm tadpole}, R}^{\mu\nu}(q^2)
=-\frac{3\alpha}\pi(-\frac 32)(2A\Lambda^4q^2)g^{\mu\nu}
\equiv -Bq^2g^{\mu\nu}.
\label{eqn:6-10}
\eeq
The total photon propagator
due to the sum of the tadpole diagrams
is given by
\beq
&&\frac {-ig^{\mu\nu}}{q^2}
+\frac {-ig^{\mu\rho}}{q^2}i\Pi^R_{{\rm tadpole},\rho\sigma}
\frac {-ig^{\sigma\nu}}{q^2}+\frac {-ig^{\mu\rho}}{q^2}
i\Pi^R_{{\rm tadpole},\rho\sigma}
\frac {-ig^{\sigma\lambda}}{q^2}
i\Pi^R_{{\rm tadpole},\lambda\tau}
\frac {-ig^{\tau\nu}}{q^2}+\cdots\nn\\[2mm]
&&=\frac {-ig^{\mu\nu}}{q^2}(1-B+B^2-+\cdots)
=\frac {-ig^{\mu\nu}}{q^2}\frac 1{1+B},
\label{eqn:6-11}
\eeq
where $B$ is defined by (\ref{eqn:6-10}).
Consequently, the tadpole contribution
(\ref{eqn:6-10}) induces a finite
charge renormalization.
Note that $B\to 0$ for $\Lambda^2a^2\to 0$.
In sharp contrast, the Lorentz-non-invariant
NCQED produces the pole-type singularity\cite{7)}
in the commutative limit
of the tadpole diagram.\footnote{The argument
in Ref.~7) goes like this.
The first integral in (\ref{eqn:6-4})
vanishes in dimensional regularization.
The second integral in (\ref{eqn:6-4})
for Lorentz-non-invariant NCQED
can easily be done in the limit $\Lambda^2\to
\infty$ for $q\circ q>0$, yielding
the result $\Pi^{\mu\nu}_{\rm tadpole}=
-(12\alpha/\pi)g^{\mu\nu}/(q\circ q)$ at
$q\circ q\to 0$. 
This pole-type
singularity is cancelled by other
one-loop diagrams with
three-point vertices, leaving an amplitude
whose commutative limit does not exist.
In our regularization scheme
we do not use dimensional regularization,
thereby retaining
the first divergent integral in (\ref{eqn:6-4}), 
which is cancelled out
by the IR divergence in the second one.}
\\
\ind
In conjunction with the gauge boson tadpole
it is instructive to
consider the tadpole diagram in Lorentz-invariant
NC $\phi^4$ model,
\beq
\hS
&=&\int\!d^{\,6}{\bar\theta}\, w({\bar\theta})
\int\!d^{\,4}x[\frac 12\partial_\mu\phi(x)*
\partial^\mu\phi(x)-\frac 12m^2\phi(x)*\phi(x)\nn\\[2mm]
&&\qquad\qquad\qquad\quad\;
-\frac {\lambda}{4!}\phi(x)*
\phi(x)*\phi(x)*\phi(x)].
\label{eqn:6-12}
\eeq
The tadpole diagram with the external momentum $p$
in this model
is given by
\beq
-i\cM_{\rm tadpole}(p^2)&=&
-\frac {i\lambda}6\int\frac{d^4l}{(2\pi)^4}
\frac i{l^2-m^2+i\epsilon}\big[
1+2\langle\cos^2{(\frac 12p\wedge l)}\rangle\big]\nn\\[2mm]
&=&-\frac {i\lambda}6
\int\frac{d^4l}{(2\pi)^4}
\frac i{l^2-m^2+i\epsilon}\big[
2+\langle\cos{(p\wedge l)}\rangle\big].
\label{eqn:6-13}
\eeq
Using the integration
(\ref{eqn:6-6}) in Euclidean metric
and converting back to Minkowski space
we have
\beq
\cM_{{\rm tadpole},
\Lambda^2}(p^2)&=&
\frac {i\lambda}3\int\frac{d^4l}{(2\pi)^4}
\frac 1{l^2-m^2+i\epsilon}|_{\Lambda^2}
+\frac \lambda{96\pi^2}
\int_0^\infty\!ds
\frac {e^{-sm^2-\frac 1{s\Lambda^2}}}
{\sqrt{s}\big(\sqrt{s-2Ap^2}\big)^3}.
\label{eqn:6-14}
\eeq
This time mass renormalization is necessary
to subtract the quadratic divergence.
The renormalized amplitude is then given by subtraction at
$p^2=m^2$:
\beq
\cM_{{\rm tadpole},\hbox{{\scriptsize ${R}$}}}(p^2)
&=&\mathop{\rm lim}_{\Lambda^2\to\infty,
a^2\to 0, \Lambda^2a^2:{\rm fixed}}\;
[\cM_{\rm tadpole, \Lambda^2}(p^2)
-\cM_{\rm tadpole, \Lambda^2}(m^2)]\nn\\[2mm]
&=&
\frac {\lambda}{96\pi^2}(-\frac 32)(2A\Lambda^4)(p^2-m^2).
\label{eqn:6-15}
\eeq
This leads to a finite wave function
renormalization. Recall that, in the usual model
defined in the commutative limit,
we have no wave function renormalization in
the one-loop approximation,
which is realized if $\Lambda^2a^2\to 0$.
\\
\ind
The fermion loop contribution
to the scalar meson self-energy
with Lorentz-invariant NC scalar coupling
\beq
G\int\!d^4xd^6\btheta w(\btheta)
{\bar\psi}(x)*\phi(x)*\psi(x)
\label{eqn:6-16}
\eeq
is given by
\beq
-i\cM_{{\rm fermion}\;{\rm loop}}(p^2)
&=&
(iG)^2(-1)\int\!\frac {d^4l}{(2\pi)^4}
{\rm Tr}
\big[
\frac i{\lslash-M+i\epsilon}
\frac i{\pslash+\lslash-M+i\epsilon}\big]\langle
e^{\frac i2p\wedge l}\rangle\langle
e^{\frac i2l\wedge p}\rangle\nn\\[2mm]
&=&
-\frac {iG^2}{4\pi^2}
\int_0^1\!dx\int_0^\infty\!dse^{-s\Delta(p^2,M^2)}\nn\\[2mm]
&&\qquad\times
\big[\frac 1{\sqrt{s}(\sqrt{s-Ap^2})^3}
-2\Delta(p^2,M^2)\frac {\sqrt{s}}{(\sqrt{s-Ap^2})^3}\big],
\label{eqn:6-17}
\eeq
where $p$ is the external momentum.
Expanding the integrands as in
(\ref{eqn:4-23}) and performing $s$-integral
with the regularization factor $e^{-\frac 1{s\Lambda^2}}$
inserted
we get
\beq
\cM_{{\rm fermion}\;{\rm loop},\Lambda^2}(p^2)
&=&
\frac {G^2}{4\pi^2}
\int_0^1\!dx
\sum_{n=0}^\infty
\left(
\ba{c}
-\frac 32\\
n\\
\ea
\right)
(Ap^2)^n\nn\\[2mm]
&&
\times\big[
2(\sqrt{\Lambda^2\Delta(p^2,M^2)})^{n+1}
K_{n+1}(2\sqrt{\Delta(p^2,M^2)/\Lambda^2})\nn\\[2mm]
&&
-2\Delta(p^2,M^2)(\sqrt{\Lambda^2\Delta(p^2,M^2)})^n
K_n(2\sqrt{\Delta(p^2,M^2)/\Lambda^2})\big].
\label{eqn:6-18}
\eeq
In the new UV limit
only $n=0,1$ terms in the first sum
and only the $n=0$ term in the second
survive.
The subtraction at $p^2=m^2$ (mass renormalization)
gives
\beq
\cM_{{\rm fermion}\;{\rm loop},\hbox{{\scriptsize ${R}$}}}(p^2,m^2)
&=&\mathop{\rm lim}_{\Lambda^2\to\infty,
a^2\to 0, \Lambda^2a^2:{\rm fixed}}\;
\big[
\cM_{{\rm fermion}\;{\rm loop},\Lambda^2}(p^2)
-\cM_{{\rm fermion}\;{\rm loop},\Lambda^2}(m^2)\big]\nn\\[2mm]
&=&\mathop{\rm lim}_{\Lambda^2\to\infty,
a^2\to 0, \Lambda^2a^2:{\rm fixed}}\;
\big[\frac {G^2}{4\pi^2}
\{(-\frac 32)(A\Lambda^4)-\frac \gamma 3\}(p^2-m^2)\nn\\[2mm]
&&\qquad\qquad\quad
+\frac {G^2}{2\pi^2}\int_0^1\!dx[\Delta(p^2,M^2)
\ln{\big(\frac{\Delta(p^2,M^2)}{\Lambda^2}\big)}\nn\\[2mm]
&&\qquad\qquad\qquad\qquad\qquad
-\Delta(m^2,M^2)\ln{\big(\frac{\Delta(m^2,M^2)}{\Lambda^2}\big)}]\big].
\label{eqn:6-19}
\eeq
The divergent coefficient, including a finite part,
of $(p^2-m^2)$
is renormalized away by
the wave function renormalization.
\section{Discussions} 
It is well-known that
$SU(N)$ cannot be employed as a gauge group
of NC Yang-Mills theory
but $U(N)$ Yang-Mills
can be formulated on NC space-time.
This is a consequence from the fact
that
the NC non-Abelian gauge field strength
\begin{eqnarray} 
F_{\mu\nu}(x)&=&\partial_\mu A_\nu(x)
-\partial_\nu A_\mu(x)
      -ie[A_\mu(x),A_\nu(x)]_*
\label{eqn:7-1}
\end{eqnarray}
has the following nonlinear term
\beq
[A_\mu(x),A_\nu(x)]_*&\equiv&
A_\mu(x)*A_\nu(x)-A_\nu(x)*A_\mu(x)\nn\\[2mm]
&=&\frac 12\sum_{a,b=0,1,\cdots,N^2-1}
(A_\mu^a(x)*A_\nu^b(x)-A_\nu^b(x)*A_\mu^a(x))
\{T_a,T_b\}\nn\\[2mm]
&&+\frac 12\sum_{a,b=1,\cdots,N^2-1}
(A_\mu^a(x)*A_\nu^b(x)+A_\nu^b(x)*A_\mu^a(x))
[T_a,T_b],
\label{eqn:7-2}
\end{eqnarray}
which contains both the commutators
and the anti-commutators
of the generators of Lie algebra.
If, however, $T_a$ denote $U(N)$ generators
as displayed in the sums of (\ref{eqn:7-2}),
they are closed under
the commutators and the anti-commutators.
In this case
the second sum in (\ref{eqn:7-2}) contains only $SU(N)$
components, whereas
the first sum involves $U(1)$
part but only through the Moyal bracket.
As in
NCQED the Moyal bracket gives rise to the Moyal
factor proportional to
$\sin{(\frac 12k_1\wedge k_2)}$ in Feynman diagrams,\footnote{Feynman
rules for Lorentz-non-invariant
NC $U(N)$ gauge theory
are collected in the Appendix of Ref.~16).}
which
vanishes upon integration
over $\theta$ as pointed out in the previous section.
This means that
$U(1)$ including ghost
decouples from the rest in three point vertices.
As a result $U(1)$ part
contributes only to the tadpole diagrams.
Tadpole diagrams containing
$U(1)$
would lead to coupling
constant renormalization.\footnote{If we further assume
the condition $\Lambda^2a^2\to 0$
in the new UV limit,
$U(1)$ 
completely decouples from the rest.}
The Moyal anti-bracket term containing
$SU(N)$ 
remains as in the commutative Yang-Mills.
This would shed light on a
consistent formulation
of $SU(N)$ gauge theory
on NC space-time.
In a separate paper we shall present a proof of
gauge invariance of NC
regularization
in Lorentz-invariant
NC pure Yang-Mills.
\\
\ind
Although we have not elaborated to
work on the Ward-Takahashi identity
for the electron vertex function involving the
electron self-energy diagram ($Z_1=Z_2$),
we have successfully formulated
a new gauge-invariant
regularization scheme starting from Lorentz-invariant NCQFT.
Anomalies are also good place to test our regularization
scheme. From our point of view
the renormalization is required to eliminate
the IR divergence in Lorentz-invariant NCQFT.
The reason that the IR divergence in Lorentz-invariant NCQFT
is connected with the UV divergence
in QFT is that
the IR limit cannot be distinguished from the
commutative limit.\footnote{Long wave length `sees' the space-time
in a coarse way, that is,
in the IR limit, the space-time non-commutativity
loses its meaning.} Lorentz invariance unravels
the fascinating aspect of the IR/UV mixing.
\\
\ind
Our use of Lorentz-invariant NCQFT
as a means of the regularization in QFT
is motivated to understand
the IR/UV mixing in an invariant way.
The elimination of the
IR singularity 
is necessitated to make
the Lorentz-invariant NCQFT useful on firm physical grounds.
There is alternative approach\cite{4),6)} to the
Lorentz-invariant NCQED using Seiberg-Witten map.\cite{11)}
It tries to look for small effects
arising from the nonvanishing small value
of the fundamental length $a$.
In this approach
Feynman rules in the theory
are the same as those of the commutative fields,
regarding the Lorentz-invariant NCQED
as an effective field theory.
There is no vertex factor like
$V(k_1,k_2)$ as introduced in \S4.
\\
\ind
Integration over $\theta$
introduces most radical noncommutativity.
It is inevitable, however, by Lorentz symmetry.
The Lorentz invariance of the average (\ref{eqn:4-9})
is proved as follows.
\bea*
V({k'}_1,{k'}_2)&=&
\langle e^{\frac i2{k'}_1\wedge {k'}_2}\rangle
=\int\,d^6{\bar\theta}w({\bar\theta})
e^{\frac i2{k'}_{1,\mu}
\theta^{\mu\nu}{k'}_{2,\nu}}\nn\\[2mm]
&=&\int\,d^6{\bar\theta}'w({\bar\theta}')
e^{\frac i2{k'}_{1,\mu}
{\theta'}^{\mu\nu}{k'}_{2,\nu}}
=\int\,d^6{\bar\theta}w({\bar\theta})
e^{\frac i2k_{1,\mu}
{\theta}^{\mu\nu}k_{2,\nu}}
=V(k_1,k_2),
\eea*
where we have used the fact that the measure
$d^6{\bar\theta}$ and the reduced weight function
$w({\bar\theta})$ are both Lorentz-invariant
and $\theta^{\mu\nu}$ is a second-rank tensor.
In this respect we
remind the readers that
the fundamental length, if any,
is reconciled with relativity only if
the notion of the continuous time-development
is abandoned.\cite{17)}
Hamiltonian formalism is no
longer tenable to derive Feynman rules with
the extra factor $V(k_1,k_2)$.
Nonetheless,
the limit (\ref{eqn:1-1})
recovers the commutative theory
as a {\it smooth} limit of Lorentz-invariant NCQFT.
\section*{Acknowledgements}
The author is grateful to
H. Kase and Y. Okumura for useful discussions.

\end{document}